\documentclass[prl,twocolumn]{revtex4}
\usepackage{graphicx}
\usepackage{amsmath}
\usepackage{epsfig}

\begin{document}

\title{Momentum-position realization of the Einstein-Podolsky-Rosen
paradox}

\author{John C. Howell$^1$, Ryan S. Bennink$^{2}$, Sean J.
Bentley$^{2*}$, R. W. Boyd$^2$}

\affiliation{$^1$Department of Physics and Astronomy, University
of Rochester, Rochester, NY 14627, USA \\ $^2$The Institute of
Optics, University of Rochester, Rochester, NY 14627, USA}

\begin{abstract}
We report on a momentum-position realization of the EPR paradox
using direct detection in the near and far fields of the photons
emitted by collinear type-II phase-matched parametric
downconversion.  Using this approach we achieved a measured
two-photon momentum-position variance product of $0.01\hbar^2$,
which dramatically violates the bounds for the EPR and
separability criteria.
\end{abstract}


\maketitle

In 1935 Einstein, Podolsky and Rosen (EPR) \cite{EPR} wrote one of
the most controversial and influential papers of the twentieth
century. They proposed a \emph{gedanken} experiment involving two
particles entangled simultaneously over a continuum of position
and momentum states. By measuring either the position or momentum
of one of the particles, the position or momentum of the other
could be inferred with complete certainty. Under the assumptions
of EPR, the ability to make such an inference meant that the
position and momentum of the unmeasured particle were simultaneous
realities, in violation of Heisenberg's uncertainty relation. This
thought experiment became known as the EPR paradox. In 1951 Bohm
\cite{Bohm} cast the EPR paradox into a simpler, discrete form
involving spin entanglement of two spin-1/2 particles, such as
those produced in the dissociation of a diatomic molecule of zero
spin. From Bohm's analysis sprang Bell's inequalities
\cite{Bell,CHSH} and much of what is now the field of discrete
quantum information
\cite{Aspect81,Ekert91,Naik00,Jennewein00,Tittel00,Bouwmeester97}.

In recent years, however, there has been a movement toward the
study of entanglement of continuous variables as originally
discussed by EPR
\cite{Franson89,Kwiat90,Rarity90,Reid88,Ou92,Duan00,Simon00,
Julsgaard01,Silberhorn01,
Mancini02,Bowen02,Korolkova02,Braunstein98,Gatti03}. Of particular
interest was the early work of Reid and Drummond \cite{Reid88}.
They derived an EPR criterion and showed how it could be
implemented with momentum-like and position-like quadrature
observables of squeezed light fields.  Shortly thereafter, the
experiment was realized by Ou \textit{et al.} \cite{Ou92}. Later
Duan \textit{et al.} \cite{Duan00} and Simon \cite{Simon00}
derived necessary and sufficient conditions for the inseparability
(entanglement) of continuous variable states.  A flurry of
experimental activity ensued in both atomic ensembles
\cite{Julsgaard01} and squeezed light fields
\cite{Silberhorn01,Bowen02,Korolkova02}.

Here we report on a demonstration of the EPR paradox using
position- and momentum-entangled photon pairs produced by
spontaneous parametric downconversion. We find that the position
and momentum correlations are strong enough to allow the position
or momentum of a photon to be inferred from that of its partner
with a product of variances $\leq 0.01\hbar^2$, which violates the
separability bound by two orders of magnitude.

In the idealized entangled state proposed by EPR,
\begin{equation}
|\mathrm{EPR}\rangle\equiv\int_{-\infty}^{\infty}|x,x\rangle\,dx=\int_{-\infty}%
^{\infty}|p,-p\rangle\, dp,\label{EPR_state}%
\end{equation}
the positions and momenta of the two particles are perfectly
correlated. This state is non-normalizable and cannot be realized
in the laboratory. However, the state of the light produced in
parametric downconversion can be made to approximate the EPR state
under suitable conditions. In parametric downconversion, a pump
photon is absorbed by a nonlinear medium and re-emitted as two
photons (conventionally called signal and idler photons), each
with approximately half the energy of the pump photon. Considering
only the transverse components, the momentum conservation of the
downconversion process requires $\mathbf{p}_{1} +\mathbf{p}_{2}
=\mathbf{p}_{p}$ where $1,2,p$ refer to the signal, idler, and
pump photons. Provided the uncertainty in the pump transverse
momentum is small, the transverse momenta of the signal and idler
photons are highly anti-correlated. The exact degree of
correlation depends on the structure of the signal+idler state. In
the regime of weak generation, this state has the form
\begin{equation}
|\psi\rangle_{1,2}=|\mathrm{vac}\rangle+\int A(\mathbf{p}_{1}%
,\mathbf{p}_{2})|\mathbf{p}_{1},\mathbf{p}_{2}\rangle
\,d\mathbf{p}_{1}d\mathbf{p}_{2}\label{PDC_state}%
\end{equation}
where $|\mathrm{vac}\rangle$ denotes the vacuum state and the
two-photon amplitude $A$ is
\begin{equation}
A(\mathbf{p}_{1},\mathbf{p}_{2})=\chi E_{p}(\mathbf{p}_{1}
+\mathbf{p}_{2})\frac{\exp(i\Delta k_{z}L)-1}{i\Delta k_{z}%
}.\label{PDC_amplitude}%
\end{equation}
Here $\chi$ is the coefficient of the nonlinear interaction,
$E_{p}$ is the amplitude of the plane-wave component of the pump
with transverse momentum $\mathbf{p}_{1}+\mathbf{p}_{2}$, $L$ is
the length of the nonlinear medium, and $\Delta
k_{z}=k_{p,z}-k_{1,z}- k_{2,z}$ (where
$\mathbf{k}=\mathbf{p}/\hbar$) is the longitudinal wavevector
mismatch, which generally increases with transverse momentum and
limits the angular spread of signal and idler photons. The vacuum
component of the state makes no contribution to photon counting
measurements and may be ignored. Also, there is no inherent
difference between different transverse components; so without
loss of generality, we consider scalar position and momentum. The
narrower the angular spectrum of the pump field and the wider the
angular spectrum of the generated light, the more closely the
integral (\ref{PDC_state}) approximates $\int_{-\infty}^{\infty}
\delta(p_{1}+p_{2})|p_{1},p_{2}\rangle\,dp_{1}
dp_{2}=|\mathrm{EPR}\rangle$ and the stronger the correlations in
position and momentum become.

The experimental setup used to determine position and momentum
correlations is portrayed in Fig.\ 1.  The idea is to measure the
positions and momenta by measuring the downconverted photons in
the near and far fields respectively \cite{Gatti03}. The source of
entangled photons is spontaneous parametric downconversion
generated by pumping a 2mm thick Type-II BBO crystal with a 30 mW,
cw, 390 nm laser beam. A prism separates the pump light from the
downconverted light. The signal and idler photons have orthogonal
polarizations and are separated by a polarizing beamsplitter. In
each arm, the light passes through a narrow 40 $\mu$m vertical
slit, a 10nm spectral filter, and a microscope objective. The
objective focuses the transmitted light onto a multimode fiber
which is coupled to an avalanche photodiode single-photon counting
module. The spectral filter ensures that only photons with nearly
equal energies are detected. To measure correlations in the
positions of the photons, a lens of focal length 100 mm (placed
prior to the beamsplitter) is used to image the exit face of the
crystal onto the planes of the two slits (Fig.\ 1a). One slit is
fixed at the location of peak signal intensity. The other slit is
mounted on a translation stage. The photon coincidence rate is
then recorded as a function of the displacement of the second
slit. To measure correlations in the transverse momenta of the
photons, the imaging lens is replaced by two lenses of focal
length 100 mm, one in each arm, a distance $f$ from the planes of
the two slits (Fig.\ 1b). These lenses map transverse momenta to
transverse positions, such that a photon with transverse momentum
$\hbar k_{\bot}$ comes to a focus at the point $x = f k_{\bot}/k$
in the plane of the slit. Again, one slit is fixed at the location
of peak count rate while the other is translated to obtain the
coincidence distribution.

\begin{figure}[tb!]
\centerline{(a)\includegraphics[scale=0.9]{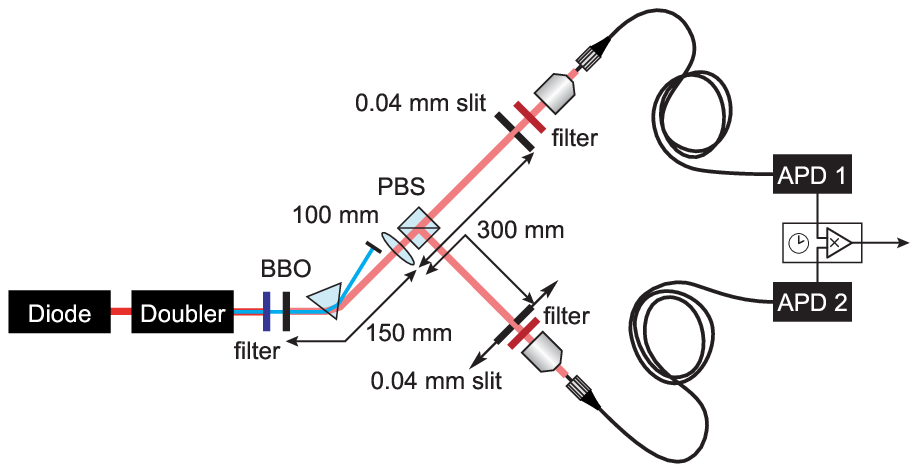}}
\centerline{(b)\includegraphics[scale=0.9]{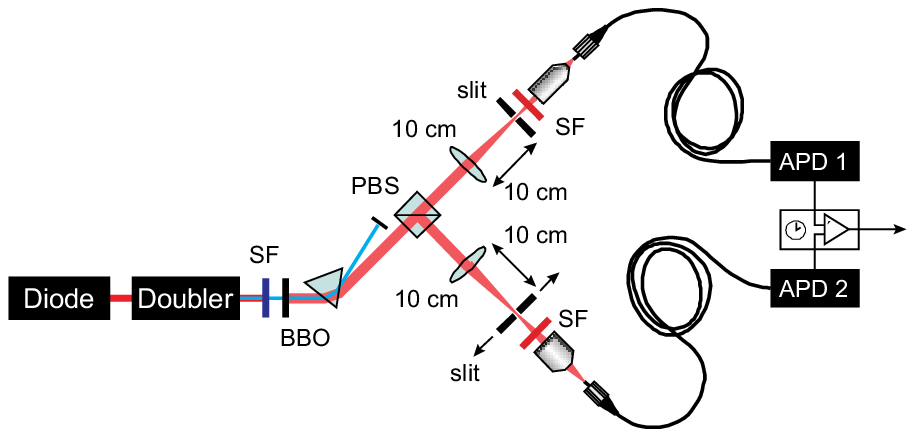}}
\caption{Experimental setup for measuring photon correlations. (a)
Position correlations are obtained by imaging the birthplace of
each photon of a pair onto a separate detector. (b) Correlations
in transverse momentum are obtained by imaging the propagation
direction of each photon of a pair onto a separate detector }
\label{Experiment}
\end{figure}

\begin{figure}[tb!]
\centerline{(a)\includegraphics[scale=0.9]{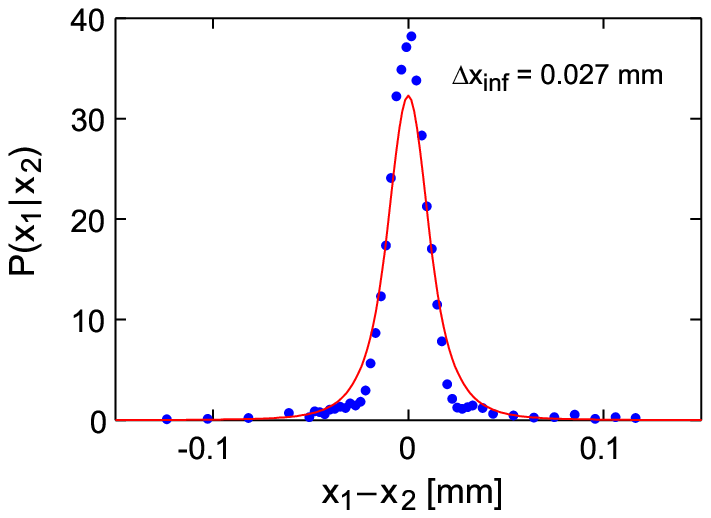}}
\centerline{(b)\includegraphics[scale=0.9]{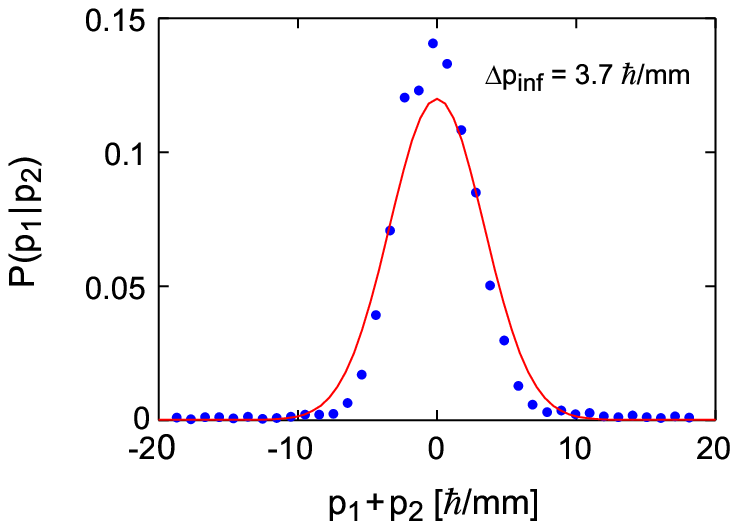}}
\caption{(a) The conditional probability distribution of relative
birthplace of the entangled photons. (b) The conditional
probability distribution of relative transverse momentum of the
entangled photons.  The widths of the distributions determine the
uncertainties in inferring the position or momentum of one photon
from that of the other.  The solid lines are the theoretical
predictions and the dots are the experimental data.}
\label{Results}
\end{figure}

By normalizing the coincidence distributions, we obtain the
conditional probability density functions $P(x_1|x_2)$ and
$P(p_1|p_2)$ (Fig.\ 2). These probability densities are then used
to calculate the uncertainty in the inferred position or momentum
of photon 1 given the position or momentum of photon 2:
\begin{align}
  \nonumber \Delta x_{\mathrm{inf}}^2 &= \int (x_1-x_2)^2 P(x_1|x_2) dx_1 \\
  & -\left( \int (x_1-x_2) P(x_1|x_2) dx_1\right)^2\\
  \nonumber \Delta p_{\mathrm{inf}}^2 &= \int (p_1+p_2)^2 P(p_1|p_2) dp_1 \\
  & -\left( \int (p_1+p_2) P(p_1|p_2) dp_1\right)^2 .
\end{align}
Because of the finite width of the slits, the raw data in Fig.\ 2
describe a slightly broader distribution than is associated with
the downconversion process itself. By adjusting the computed
values of $\Delta x_{\mathrm{inf}}$ and $\Delta p_{\mathrm{inf}}$
to account for this broadening (an adjustment smaller than 10\%),
we obtain the correlation uncertainties $\Delta x_{\mathrm{inf}} =
0.027\operatorname{mm}$ and $\Delta p_{\mathrm{inf}} = 3.7
\hbar\operatorname{\,mm}^{-1}$. The measured variance product of
the inferred state is
\begin{equation}
\Delta x_{\mathrm{inf}}^2\Delta p_{\mathrm{inf}}^2 = 0.01\hbar^2.
\end{equation}

Also shown in Fig.\ 2 are the predicted probability densities.
These curves contain no free parameters and are obtained directly
from the two-photon amplitude $A(p_1,p_2)$, which is determined by
the optical properties of BBO and the measured profile of the pump
beam. Fig.\ 2 indicates that the correlation widths we obtained
are intrinsic to the downconversion process and are limited only
by the degree to which it deviates from the idealized EPR state
(\ref{EPR_state}). The value of $\Delta p_\mathrm{inf}$ is limited
by the finite width of the pump beam. The pump photons in a
Gaussian beam of width $w$ have an uncertainty $\hbar/2w$ in
transverse momentum which, due to conservation of momentum, is
imparted to the total momentum $p_1+p_2$ of the signal and idler
photons. The value of $\Delta x_\mathrm{inf}$ is limited by the
range of angles over which the crystal generates signal and idler
photons. If the angular width of emission is $\Delta\phi$, then
the principle of diffraction indicates that the photons cannot
have a smaller transverse dimension than $\sim
(k_{s,i}\Delta\phi)^{-1}$. Careful analysis based on the angular
distribution of emission yields $\Delta x_\mathrm{inf} =
1.88(k_{s,i}\Delta\phi)^{-1}$. With the measured beam width of
$w=0.17\operatorname{mm}$ and predicted angular width
$0.012\operatorname{rad}$, the theory predicts $\Delta
x_\mathrm{inf}^2 \Delta p_\mathrm{inf}^2 = 0.0036\hbar^2.$ This is
somewhat smaller than the experimentally calculated value of
$0.01\hbar^2$, even though the data appears to closely match the
theoretical curves. The reason for this discrepancy is that the
experimental distributions have small ($\approx 1$\% of the peak)
but very broad wings. The origin of these coincidence counts is
unknown; they are perhaps due to scattering from optical
components. If these counts are treated as a noise background and
subtracted, the experimentally obtained uncertainties come into
agreement with the theoretically predicted values, yielding an
uncertainty product of $0.004\hbar^2$.

To interpret these results, it is helpful to consider the
relationship between the original EPR paradox and the issues of
entanglement, non-locality, and quantum signatures, which have
been the subjects of more modern studies. Because of lingering
doubts about quantum theory, raised in part by the EPR paradox,
there has been much interest in ``quantum signatures'', i.e.\
phenomena that confirm the predictions of quantum mechanics and
cannot be explained by classical mechanics. However, the intent of
EPR was not to reveal a discrepancy between classical and quantum
theory, but to reveal an inconsistency, or incompleteness, within
quantum theory. This inconsistency was revealed by showing that
measurements of one particle could be used (seemingly) to infer
the state of an unmeasured particle with greater certainty than is
allowed by the uncertainty relation, i.e. $\Delta
x_{\mathrm{inf}}^2\Delta p_{\mathrm{inf}}^2\geq\hbar^2/4$
\cite{EPR,Reid88}. A key component of EPR's argument was the
assumption of ``local reality.'' Under this assumption, the
statistics of each particle depend only on a (hidden) state
parameter $s$ which is determined while the particles are close
enough to interact. This assumption is commonly taken to mean that
the joint probablity of any pair of observables $a_{1}$ and
$b_{2}$ must be expressible as
\begin{equation}
P(a_{1},b_{2})=\sum_{s}P_{s}P(a_{1}|s)P(b_{2}|s)\label{locality_criterion}
\end{equation}
where $P_{s}$ is the probability of state $s$ and $P(a_{1}|s)$ and
$P(b_{2}|s)$ are the conditional probabilities for particle 1 and
particle 2 respectively. This constraint led to the Bell tests of
local realism \cite{Bell,CHSH} and their subsequent experimental
realizations (e.g., \cite{Aspect81}). Another consequence of
\cite{EPR} has been the development of the concept of
entanglement, which is closely connected to non-locality. A
bipartite system is entangled (inseparable) if its density matrix
\emph{cannot} be written in the form
\begin{equation}
\rho=\sum_{s}P_{s}\rho_{1,s}\otimes\rho_{2,s}.\label{separability_criterion}%
\end{equation}
Measurement of either member of an entangled system projects both
members into a mixture of states consistent with the result of the
measurement. It is now generally accepted that this mutual
projection occurs even when the particles are widely separated;
hence, within the framework of quantum theory, entanglement and
non-local behavior have a one-to-one relationship. This conclusion
is confirmed by the fact that any system which satisfies eqn.
(\ref{separability_criterion}) also satisfies eqn.
(\ref{locality_criterion}), and vice versa. However, entanglement
does not always rule out local realism in the context of a theory
other than quantum mechanics. For example, the strong position and
momentum correlations of the entangled EPR state can be readily
explained by local-realistic classical mechanics, which does not
impose an uncertainty relation on position and momentum.

The close connection between entanglement and non-locality has
prompted refinement of EPR's criterion, resulting in a number of
different separability tests and entanglement measures. Of
relevance to the original EPR paradox are the tests for
separability of continuous- variable systems developed by Duan
\emph{et al.} \cite{Duan00}, Simon \cite{Simon00}, and Mancini
\emph{et al.} \cite{Mancini02}. The tests in \cite{Duan00} and
\cite{Simon00} involve sums of dimensionless variances and are not
scale invariant; hence it is not clear how, or even if, they may
be applied to the present EPR experiment, which involves
dimensional position and momentum. The criterion derived by
Mancini \emph{et al.} \cite{Mancini02} is more useful here. It
states that separable systems must satisfy the joint uncertainty
relation
\begin{equation} (\Delta x_{12})^2(\Delta
p_{12})^2\geq\hbar^2\label{Mancini_criterion}%
\end{equation}
where $(\Delta
x_{12})^2=\langle(x_{1}-x_{2})^2\rangle-\langle(x_{1}-x_{2})\rangle^2$
and $(\Delta
p_{12})^2=\langle(p_{1}+p_{2})^2\rangle-\langle(p_{1}+p_{2})\rangle^2.$
The angle brackets denote expected values over the respective
joint probability distributions.  In our experiment the widths of
the conditional probability distributions $P(x_1|x_2)$ and
$P(p_1|p_2)$ are essentially independent of $x_2$ and $p_2$ over
most of their ranges, so that $\Delta x_\mathrm{inf}^2$ and
$\Delta p_\mathrm{inf}^2$ are nearly equal to $(\Delta x_{12})^2$
and $(\Delta p_{12})^2$. Hence our results constitute a two
order-of-magnitude violation of the joint uncertainty relation
(\ref{Mancini_criterion}), which in this case is both a
separability criterion and a local-realism criterion.

Finally, we note that the EPR paradox does not represent a true
inconsistency. It is generally accepted that the EPR argument
fails because the assumption of local realism is invalid: The
position or momentum of the unmeasured particle becomes a reality
when, and only when, the corresponding quantity of the other
particle is measured. As the measurement involves only one
quantity or the other, the position and momentum of the unmeasured
particle need not be simultaneous realities.

In conclusion, we have reported the experimental realization of
Einstein, Podolsky and Rosen's paradox using momentum-position
entangled photons. We have measured position and momentum
correlations resulting in a variance product which dramatically
violates the original EPR criterion and a modern inseparability
criterion. Compared to squeezed-light realizations of the EPR
paradox, the momentum-position realization has several attractive
features which make it promising for further development. For one,
the entanglement is observed using direct photon detection, which
is experimentally simpler than homodyne detection. Secondly, the
entanglement does not reside in the photon count, which frees this
quantity to be used for postselection. Since the position and
momentum measurements involve only those photons that are
detected, the measured entanglement is not degraded by optical
loss which inevitably occurs in real systems. For both of these
reasons, systems with very small values of the joint uncertainty
product can be readily achieved in practice. This capability has
already been used to achieve near- diffractionless coincidence
imaging \cite{Bennink}. We believe that the work presented here
sets the stage for many more interesting applications to come.

JCH acknowledges support from the NSF and University of Rochester.
RWB gratefully acknowlwedges  support  by ARO under award
DAAD19-01-1-0623 and ONR under award N00014-02-1-0797

$^*$ Sean Bentley is now at Adelphi University.

\end{document}